\documentstyle[11pt,psfig]{article}
\def\reference{\parskip 0pt\par\noindent\hangindent 0.5 truecm}

\begin{document}
%
%
\title{Extreme emission line outflows in the GPS source \\4C 12.50
(PKS 1345+12)}
%


\author{J. Holt, $^{1}$ 
 C. N. Tadhunter, $^{1}$ and
 R. Morganti $^{2}$
} 

\date{ }
\maketitle

{\center
$^1$ The University of Sheffield, Hicks Building, Hounsfield Road,  
Sheffield, S3 7RH, UK.\\j.holt@sheffield.ac.uk, c.tadhunter@sheffield.ac.uk\\[3mm]
$^2$ ASTRON, PO Box 2, 7990 AA Dwingeloo, The
Netherlands.\\morganti@astron.nl\\[3mm] 
}

\begin{abstract}
We present high resolution spectra (0.7 $\AA$/pix) of the GPS source
4C 12.50 with large spectral coverage ($\sim 4500\AA$) taken with the
4.2m William Herschel Telescope, La Palma.  
The slit was aligned along PA 160 to include the
nucleus and emission line region to the NW. An asymmetric halo
extending 20 kpc NW and 12 kpc SE from the nucleus is clearly
seen. At the position of the nucleus we observe unusually broad
forbidden emission line components (broadest component: FWHM $\sim 2000$ km s$^{-1}$), blue shifted
by up to 2000 km s$^{-1}$ with respect to the halo of the galaxy and
HI absorption. We interpret this as material in outflow. 
We measure E(B-V) = 1.44 for the broadest, most kinematically
disturbed  component, corresponding to an
actual H$\beta$ flux 130 times brighter than that measured. We
calculate an upper limit for the mass of the line emitting gas of
order 10$^{6}$ M$_{\odot}$ for both the intermediate and broad
components. Our results are consistent with 4C 12.50 being a young
radio source.
\end{abstract}

\vspace*{5pt}
\section{Introduction}
The narrow line region (NLR $<$ 5 kpc) is an important probe of the
ISM close to the central engine in radio galaxies (Tadhunter {\it{et
al.}}, 2001). 

Powerful {\it{extended}} radio galaxies have kinematics generally
consistent with gravitational motions in the bulges of the host
galaxies (e.g. Heckman {\it{et al.}}, 1984). In contrast, GPS radio
sources ($<$ 1 kpc in size) show evidence for kinematic disturbance
(Gelderman \& Whittle, 1994). Highly complex broad emission lines
are observed but the nature of the emission line kinematics is still
not understood. Asymmetries are believed to be the spectral signatures
of systematic flows, but their direction (outflow or inflow) is
still under dispute due to the lack of understanding of the
dust distribution in the NLR, and lack of accurate host galaxy (rest
frame) redshifts.

Compact radio sources are thought to be either young radio sources
(e.g. Fanti {\it{et al.}}, 1995) or intrinsically compact objects
confined by an unusually dense ISM (the `frustration theory', e.g. van
Breugel, 1984). High reddening is expected in both scenarios - the
nucleus is likely to be enshrouded by large amounts of gas and dust
left over from the event(s) which triggered the activity (e.g. a
merger, Heckman {\it{et al.}}, 1986). With evolution, material along
the radio axis will be dissipated by jet-cloud interactions and/or
quasar-induced winds (Tadhunter {\it{et al.}}, 2001). Eventually
cavities will be hollowed out on either side of the nucleus, as is seen in
the powerful extended radio source Cygnus A (Tadhunter {\it{et al.}}, 1999).

 \begin{figure}[t]
 \centerline{\psfig{file=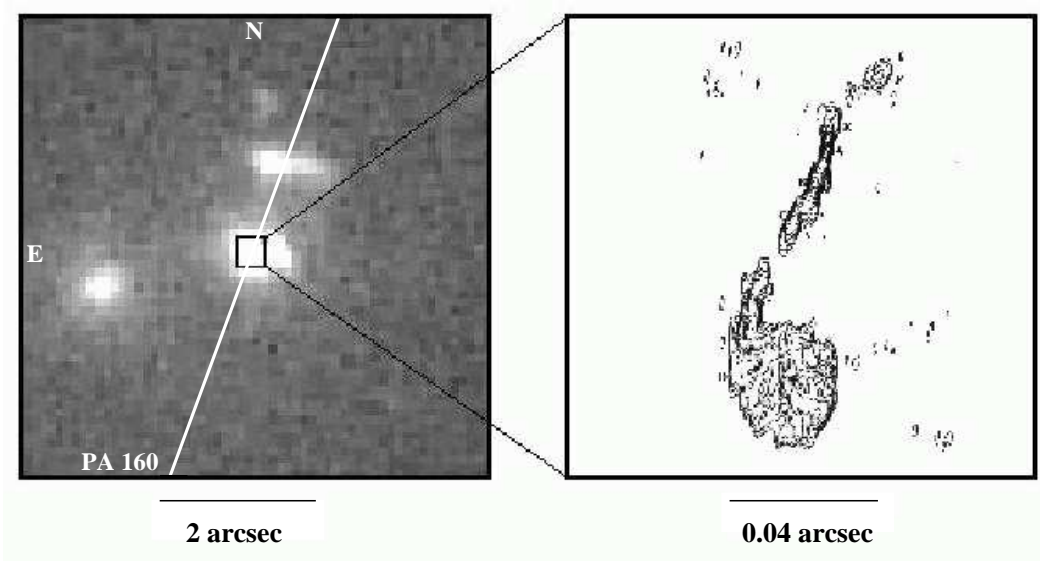,width=14cm}}
 \caption{High contrast HST WFPC2 image in [O III] from Axon {\it{et
 al.}} (2000) with the VLA radio map (5 GHz) from Stanghellini {\it{et al.}}
 (1997). For a higher resolution map, refer to the original publication.}
 \label{Figure 1}            
 \end{figure} \vspace*{5pt}

\begin{figure}[t]
\centerline{\psfig{file=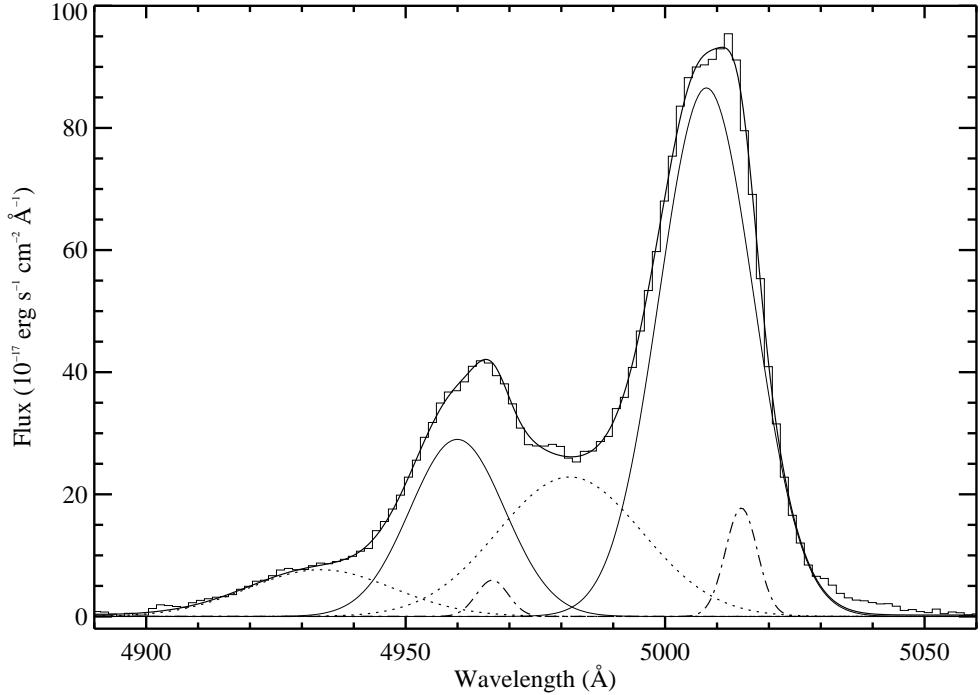,width=14cm}}
\caption{Model for [O III]$\lambda\lambda$4959,5007 (bold line)
comprising 3 components for each line: narrow (dot-dashed line), intermediate
(solid line) and broad (dotted line).}
\label{Figure 2}
\end{figure}

\section{Previous observations}
4C 12.50 has a double nucleus and shows clear evidence for
 distorted morphology in the ionised gas: high resolution HST images reveal 
an arc-like 
 structure 1.2 arcsec (2.8 kpc) N of the nucleus 
and fainter diffuse emission 2
 arcsec (4.7 kpc) N of the nucleus (see figure 1). The radio source has
 distorted, triple morphology aligned along PA 160 with 
 emission confined to a region $<$ 0.1 arcsec ($\sim$ 240 pc) in
 size (see figure 1). A bright knot towards the northern limit of the jet is
 tentatively identified as the core (Stanghellini {\it{et al.}},
 1997). The jet extends 0.04 arcsec (95 pc) to the SE before bending
 and expanding into a diffuse lobe. Weak emission extends up to 0.03
 arcsec (70 pc) NW of the core. (H$_0$ = 75 km s$^{-1}$, q$_0$ = 0.0).

\section{Results}
These {\em preliminary} results are based on new, high resolution spectroscopic
observations ($\sim
0.7\AA$/pix) with large spectral coverage ($\sim 4500\AA$) taken in
2001 May with
the 4.2m William Herschel Telescope (WHT) on La Palma. For
all observations, the slit width was 1.3 arcsec and the seeing was
$\sim$ 1.1 arcsec (FWHM) or better. For a complete analysis and
discussion of PKS 1345+12, see Holt {\it et al.} (2002, in preparation).

\subsection{Modelling the emission
lines}
The emission lines were modelled using the Starlink package {\sc
DIPSO} constraining, for example, the intensity ratios and shifts
between doublet lines, where possible, in accordance with atomic physics. 
In order to model the unusually broad [O III]$\lambda\lambda$4959,5007
emission lines in the nucleus 
it is essential to use three Gaussians (see figure 2).
These three components represent line emission from three distinct
redshift systems: \pagebreak \\
\noindent {\it{i)}} A narrow component (FWHM $\sim$ 360 km s$^{-1}$) assumed
  to emanate from the ambient, quiescent ISM. Its velocity is
  consistent with the deep HI 21cm absorption (Mirabel, 1989; see also 
Morganti
  {\it{et al.}}, these proceedings), and the extended [O II] halo.\vspace*{5pt}  \\
{\it{ii)}} An intermediate component (FWHM $\sim$ 1260 km s$^{-1}$)
blue-shifted by $\sim$ 400 km s$^{-1}$ relative to the narrow
component. \vspace*{5pt} \\
{\it{iii)}} A broad component (FWHM $\sim$ 1960 km s$^{-1}$) blue-shifted
by $\sim$ 1980 km s$^{-1}$ relative to the narrow component.\\

\begin{figure}[t]
\centerline{\psfig{file=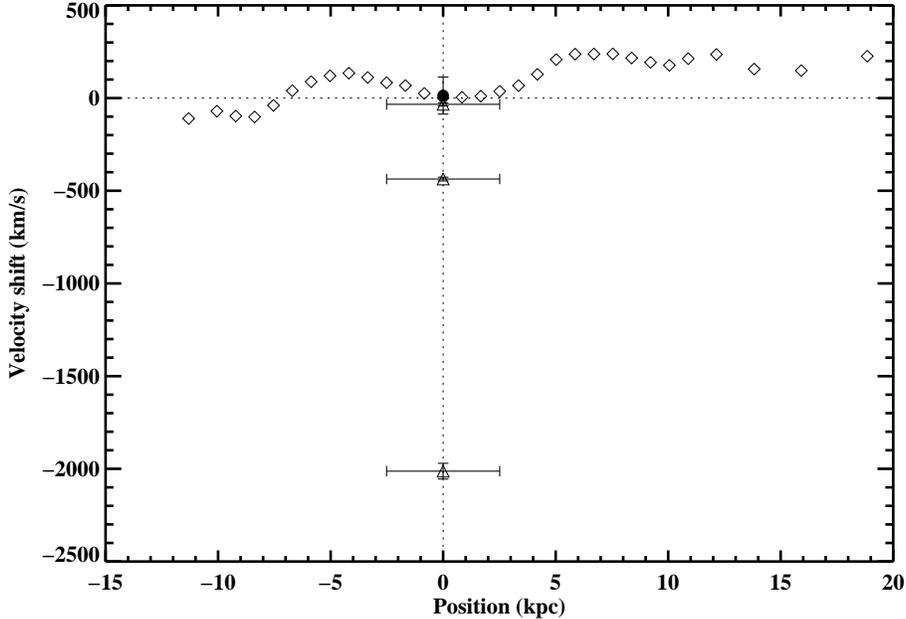,width=12.5cm}}
\caption{Comparison of the three components of [O III] (open triangles) with the spatially extended [O II]$\lambda\lambda$3726,3729 emission (open diamonds) and the deep HI 21cm absorption from Mirabel {\it et al.} (1989) (filled circle) provides a good estimate of the galaxy rest frame. Two of the [O III] components trace material in outflow. }
\label{Figure 3}
\end{figure}

The velocity shifts in the [O III] lines clearly indicate that they
originate from three distinct regions along the
line-of-sight. The narrowest component of [O III] is consistent with
the spatially extended [O II] emission and HI absorption and
represents the galaxy rest frame.
Hence, the intermediate and broad components of [O III] trace material
blue-shifted with respect to the galaxy system, suggesting that the
ISM emitting these components is in outflow (see figure 3).

\subsection{Evidence for high reddening in the nucleus}
We measure E(B-V) values, 
calculated from the H$\alpha$/H$\beta$ ratio, of 0.06, 0.42 and 1.44
for the narrow, intermediate and broad components respectively. The
broad component is highly reddened and corrections for reddening yield
H$\beta$ fluxes a factor of 130 brighter than those measured. This
reddening is consistent with Pa$\alpha$/H$\alpha$ ratios estimated
using Pa$\alpha$ measurements from Veilleux {\it{et al.}}
(1997). Note, the highly broadened Pa$\alpha$ component detected by
Veilleux {\it{et al.}} is not associated with a classical Broad Line
Region (BLR) as it is also observed in the forbidden lines and
therefore represents a broadened NLR component.

\subsection{Estimating the gas mass}
Emission line luminosities are related to the mass of line emitting
gas by: \\

\hspace*{5cm} M$_{gas}$ = m$_p$  $\frac{L(H\beta)}{N_e
\alpha_{H\beta}^{eff} h \nu_{H\beta} }$ \hfill \vspace*{12pt}\\
where N$_e$ is the electron density
(cm$^{-3}$); m$_p$ is the mass of the proton (kg); 
L(H$\beta$) is the luminosity of the
H$\beta$ line: 0.44 $\pm$ 0.04, 5.50 $\pm$ 0.20 and 26.59 $\pm$ 0.59
(10$^{40}$ erg s$^{-1}$) for the narrow, intermediate and broad
components respectively; $\alpha_{H\beta}^{eff}$ is the effective
recombination coefficient for H$\beta$ (cm$^{3}$ s$^{-1}$; see
Osterbrock 1989) and 
h$\nu_{H\beta}$ is the energy of the H$\beta$ photons (erg). 

The density diagnostic, the [S II]$\lambda\lambda$6716,6730 doublet,
can only be modelled with very high densities for the intermediate and
broad components. We estimate lower limits for the density of
the intermediate and broad components of $\sim
4000$ cm$^{-3}$. These lead to an upper limit for the total mass
of the line emitting
gas of 9.17 $\times 10^5$ M$_{\odot}$. 

 \begin{figure}[t]
 \centerline{\psfig{file=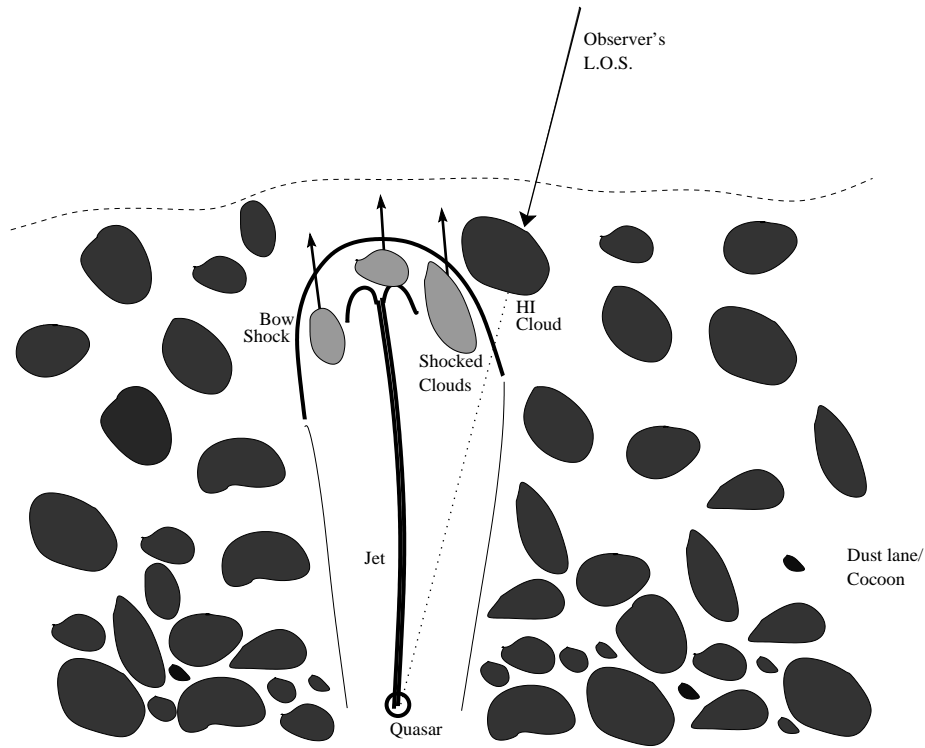,width=12.2cm,angle=-90.}}
 \caption{Model proposed for the compact flat spectrum radio source PKS 1549-79 by Tadhunter {\it{et al.}} (2001).}
 \label{Figure 4}            
 \end{figure}

\section{Conclusions} 
Our data show:\\
$\bullet$ complex broad emission lines;\\
$\bullet$ blue shifts up to $\sim$ 2000 km s$^{-1}$ with respect
to the galaxy halo and HI absorption;\\
$\bullet$ large reddening (broadest component E(B-V) $\sim 1.44$);\\
$\bullet$ high densities and high ionisation for the intermediate and
broad components. \vspace*{12pt}\\
For the blue shifted components we calculate:\\
$\bullet$ a lower limit for the density;\\
$\bullet$ an upper limit for the gas mass of order 10$^6$ M$_{\odot}$. 
\pagebreak \\
Our results are consistent with the model proposed for the compact
flat spectrum radio source PKS 1549-79 by Tadhunter {\it{et al.}}
(2001) in which the nucleus is cocooned in gas and dust in the early
stages and drives outflows which will eventually hollow out the
ionisation cones on either side of the nucleus (see figure 4).

Our estimates for the mass of emitting gas of order 10$^6$ M$_{\odot}$
are also consistent with the radio source being young. Using the
equations presented by O'Dea (1998) from de Young (1993), a smooth
medium with the density we measure would require gas masses of order
10$^8$ M$_{\odot}$ to 10$^{10}$ M$_{\odot}$ to frustrate the radio
source - much larger than we estimate for the blue shifted components. 


\section*{Acknowledgements}
\small
JH acknowledges a PPARC PhD studentship. 
The William Herschel Telescope is operated on the
island of La Palma by the Isaac Newton Group in the Spanish
Observatorio del Roque de los Muchachos of the Instituto de
Astrofisica de Canarias. 

\scriptsize
\section*{References}
\reference Axon D. J., Capetti A., Fanti R., Morganti R., Robinson
A. \& Spencer R., 2000, AJ, 120, 2284
\reference de Young D. S., 1993, ApJ, 402, 95
\reference Fanti C., Fanti R., Schilizzi R. T., Spencer R. E. \&
Stanghellini C., 1995, A\&A, 302, 317
\reference Gelderman R. \& Whittle M., 1994, ApJS, 91, 491
\reference Heckman T. M., Miley G. K. \& Green R. F., 1984, ApJ, 281,
525
\reference Heckman T. M., Smith E. P., Baum S. A., van Breugel W. J. M., Miley G. K., Illingworth G. D., Bothun G. D. \& Balick B., 1986, ApJ, 311, 526
\reference Mirabel I. F. et al, 1989, ApJ, 340, L13
\reference O'Dea, C. P., 1998, PASP, 110, 493
\reference Osterbrock D. E., 1989, {\it Astrophysics of Gaseous Nebulae
  and Active Galactic Nuclei}, University Science Books, p. 80
\reference Stanghellini C., O'Dea C. P., Baum S. A., Dallacasa D.,
Fanti R. \& Fanti C., 1997, A\&A, 325, 943
\reference Tadhunter C. N., Packham C., Axon D. J., Robinson A., Hough J., Young S., Sparks W., 1999, ApJ, 512, L91
\reference Tadhunter C. N., Wills K., Morganti R., Oosterloo T. \&
Dickson R., 2001, MNRAS, 327, 227
\reference van Breugel W. J. M., 1984, IAU symp., 110, p.59
\reference Veilleux S., Sanders D. B. \& Kim D.-C., 1997, ApJ, 484, 92

\end{document}